\documentclass[a4paper]{article}

\usepackage{INTERSPEECH2022}
\usepackage{algorithm}
\usepackage[noend]{algpseudocode}
\usepackage{amssymb}
\usepackage{xcolor}
\usepackage{url}
\usepackage{multirow}
\usepackage{booktabs}
\usepackage{wrapfig}
\usepackage{kotex}

\title{Domain Generalization with Relaxed Instance Frequency-wise Normalization for Multi-device Acoustic Scene Classification}
\name{Byeonggeun Kim, Seunghan Yang, Jangho Kim$^{*}$, Hyunsin Park, Juntae Lee, \& Simyung Chang}
\address{
Qualcomm AI Research${}^{\dagger}$, Qualcomm Korea YH, Seoul, Republic of Korea  \thanks{  ${}^{\dagger}$ Qualcomm AI Research is an initiative of Qualcomm Technologies, Inc. ${}^{*}$Author completed the research in part during an internship at Qualcomm Technologies, Inc.}
}
\email{\{kbungkun, seunghan, jangkim, hyunsinp, juntlee, simychan\}@qti.qualcomm.com\\}

\begin{document}

\maketitle
\begin{abstract}
  While using two-dimensional convolutional neural networks (2D-CNNs) in image processing, it is possible to manipulate domain information using channel statistics, and instance normalization has been a promising way to get domain-invariant features. 
  However, unlike image processing, we analyze that domain-relevant information in an audio feature is dominant in frequency statistics rather than channel statistics. Motivated by our analysis, we introduce \textit{Relaxed Instance Frequency-wise Normalization} (RFN): a plug-and-play, explicit normalization module along the frequency axis which can eliminate instance-specific domain discrepancy in an audio feature while \textit{relaxing} undesirable loss of useful discriminative information. Empirically, simply adding RFN to networks shows clear margins compared to previous domain generalization approaches on acoustic scene classification and yields improved robustness for multiple audio devices. Especially, the proposed RFN won the DCASE2021 challenge TASK1A, low-complexity acoustic scene classification with multiple devices, with a clear margin, and RFN is an extended work of our technical report \cite{resnorm_dcase_challenge}.
\end{abstract}

\section{Introduction}

Deep neural networks (DNNs) have difficulty being generalized to unseen domains, which can cause poor results in real-world scenarios. Hence, in most fields, including computer vision, audio processing, and natural language processing, \textit{domain generalization} (DG) has been an essential research topic.

Since \cite{lecun_firstconv_paper} and \cite{alexnet} introduced two-dimensional convolutional neural networks (2D-CNNs) in the image recognition task, 2D-CNN architecture has been widely employed beyond the image field. Attention to that domain-relevant information is reflected in the channel statistics of the convolutional features of images; several works~\cite{BIN, IBNNet, metabin, tasknorm} exploited Instance Normalization (IN)~\cite{IN, IN2} to eliminate instance-specific domain discrepancy. 
Here, we raise a question: \textit{Does this approach make sense in other fields, especially in audio?}

In audio, both temporal and frequency dimensions convey essential information. Thus, it has been \textit{de facto} to represent audio signals with 2D representation such as log-Mel spectrogram and Mel-Frequency Cepstral Coefficients (MFCCs). Like the image field, taking those 2D representations as inputs, 2D-CNNs have been adopted in diverse audio tasks, e.g., audio scene classification, speech recognition, and speaker recognition. However, unlike images where 2D convolution is operated along spatial dimensions, 2D convolution operates on the frequency and temporal information in the audio field. Hence, domain information may not be mainly distributed in channel statistics in audio. Although various works have addressed better usage of 2D-CNNs in the audio field~\cite{CNNinAudio, specaugment, subspectralnet, conv_freqhighlow, rethink_cnn_foraudio, bcresnet}, there has been less effort to focus on dimensional characteristics of 2D audio representation for DG. 

This work focuses on DG with explicit normalization, manipulating statistics in 2D audio features. In particular, we analyze the relationship between the domain and the statistics of each feature dimension by estimating mutual information. It demonstrates that the \textit{frequency feature} (including spectrum and cepstrum) dimension carries more domain-relevant information than the channel dimension. Moreover, we find that the frequency dimension also contains a meaningful representation of class discriminative information. Motivated by the analysis, we introduce a plug-and-play DG module named Relaxed instance Frequency-wise Normalization (RFN). RFN removes domain information through normalization along frequency dimension at the input and hidden layers of networks. We also introduce the term \textit{relaxation} that can effectively control the degree of normalization to prevent the loss of helpful class discriminative information. We provide analysis of RFN on a multi-device acoustic scene classification (ASC) task, which is challenging due to the domain gap raised by multiple devices. We empirically observe that applying DG techniques to the frequency dimension benefits audio features in 2D-CNNs. Furthermore, we show that RFN can be extended to other audio tasks, i.e., keyword spotting (KWS) and speaker verification (SV).

\section{Characteristics of Audio in 2D CNNs}
\label{sec:2d_audio}
\subsection{Preliminaries}
\noindent \textbf{Notations.} We consider 2D audio representations in $R^{F\times T}$ as an input of 2D-CNNs, where $F$ and $T$ stand for frequency and time axes, respectively. With a mini-batch axis $N$, we can represent an element of activation, $x$, by a 4D-index of $i = (i_N,i_C,i_F,i_T)$, where $i_C$ is a channel-index, where an activation, $x_{i_N, :, :, :}$ (or simply $x_{i_N}$) $\in R^{1\times C\times F\times T}$. We utilize instance statistics, mean and standard deviation (std), along a specific dimension. For example, we denote instance frequency-wise statistics of $x_{i_N}$ as $s^{(F)}$ (omit index $i_N$ for clarity) which is a concatenation of mean and std along $F$-axis, $s^{(F)} = \text{concat}(\mu^{(F)}, \sigma^{(F)})$, where $\mu^{(F)}, \sigma^{(F)} \in R^F$. 
There are spatial axes, $H$ (height) and $W$ (width) instead of $F$ and $T$ in images. Motivated by that the instance statistics along a specific dimension can represent domain-relevant information~\cite{BIN, adain, DG_mixstyle}, we analyze the characteristics of each dimension in 2D-CNNs.

\begin{figure*}[t]
  \centering
  \includegraphics[width=0.95\linewidth]{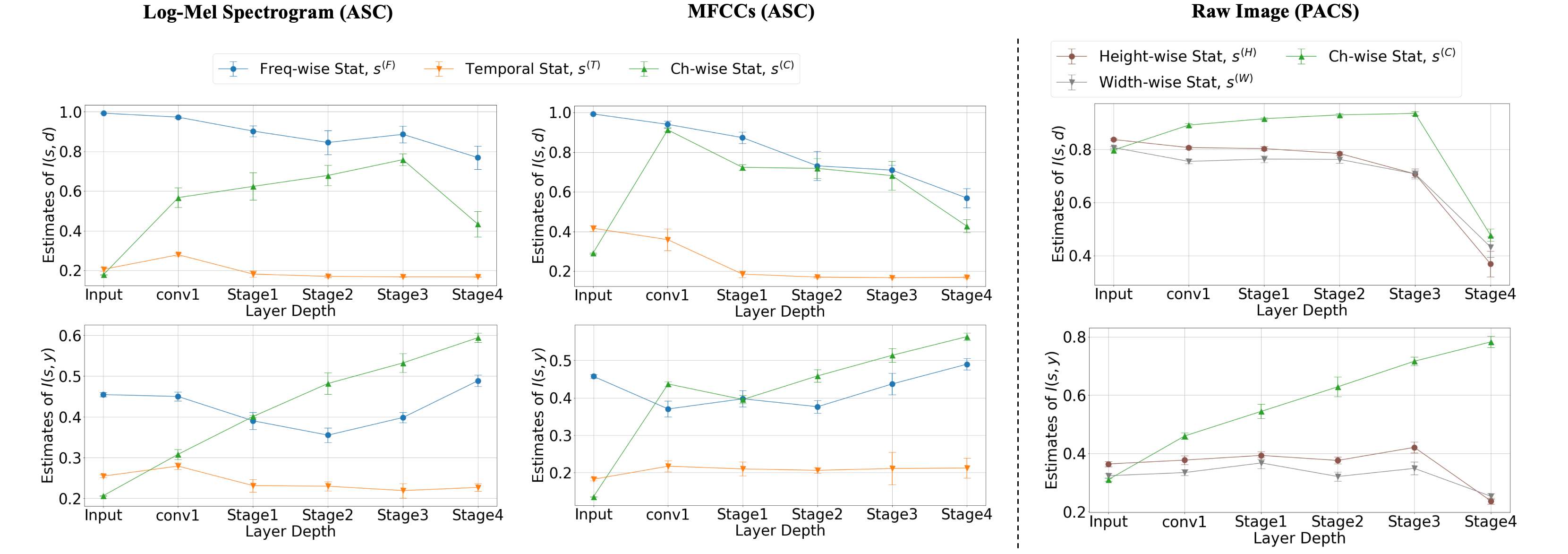}
  \vskip -0.13in
  \caption{\textbf{Estimates of mutual information} between dimension-wise statistics of intermediate hidden feature and the domain label, $I(s, d)$ (top), or the class label, $I(s, y)$ (bottom). We use log-Mel spectrogram (left) and MFCC (middle) on the audio scene classification task and use raw RGB image (right) for the PACS dataset (Average over five seeds; error bar stands for standard deviation). 
  }
  \vskip -0.2in
  \label{fig:mutual_info}
\end{figure*}

\noindent \textbf{Estimation of mutual information.} Direct computing of mutual information (MI) is usually intractable for continuous random variables. Following the literature~\cite{get_mutual_info}, we add an auxiliary classifier with parameter $\psi$ on top of the statistics along a specific dimension of a hidden activation $x$. We train the classifier to correctly classify $d$ or $y$, indicating one-hot ground-truth domain or task label, respectively. In detail, MI is denoted by $I(x, y)=H(y)-E_{p(x, y)}[-\log p(y|x)]$. We approximate the expectations of true $p(y|x)$ as the mean of $q_{\psi}(y|x)$ over a training set of size $M$. The resulting estimation of $I(x, y)$ is $H(y) - \frac{1}{M} \sum^M_{i=1}-\log q_{\psi}(y_i|x_i)$. \cite{get_mutual_info} simply uses the test accuracy of $q_{\psi}(y|x)$ as an estimation of $I(x, y)$, which is highly correlated to $\frac{1}{M} \sum^M_{i=1}-\log q_{\psi}(y_i|x_i)$, the cross-entropy loss (same for $d$).

\subsection{Analysis of Instance Statistics of 2D Audio Features}

For the MI analysis in images, we employ ResNet18~\cite{resnet} trained on a multi-domain image dataset, PACS~\cite{pacs_dataset}, which has seven common categories for $y$ with four domains (photo, art-painting, cartoon, and sketch) for $d$. In audio, we use BC-ResNet-Mod-1~\cite{resnorm_dcase_challenge, bcresnet} (denote as BC-ResNet for simplicity) suitable for DCASE 2021, multi-device acoustic scene classification (ASC) task~\cite{dcase_dataset} which has ten acoustic scenes (e.g., airport and shopping mall) for $y$ with six train-recording devices (domains) for $d$. 
We did a six-way classification for seen devices (3 real and 3 simulated) to estimate $I(s, d)$, using two representative audio features: log-Mel spectrogram and MFCCs. We estimate MI at the input and after each \textit{stage}, a sequence of convolutional blocks whose activations have the same width.

The first row of Figure~\ref{fig:mutual_info} shows the estimates of MI of each dimension with varying stages for domain labels. In PACS (right), the estimate of $I(s^{(C)}, d)$ is higher than others in every stage.
The observation correlates well with the common belief that channel-wise instance statistics can represent \textit{style} (domain) in image-level 2D-CNNs~\cite{adain}. Contrarily, in audio (left and middle), MI of frequency-wise statistics (Freq-wise Stat) is a par or superior to that of the channel and temporal statistics.
The second row of Figure~\ref{fig:mutual_info} shows the estimates of MI for task label $y$. In the image domain, channel statistics yield more dominant values than the width or height-wise ones. On the other hand, frequency statistics are also highly correlated to class label $y$ in the audio domain. Therefore, we can infer that it is essential to suppress unnecessary domain information while preserving task-relevant information in the frequency dimension in audio.

\begin{figure}[t]
  \centering
  \includegraphics[width=0.95\linewidth]{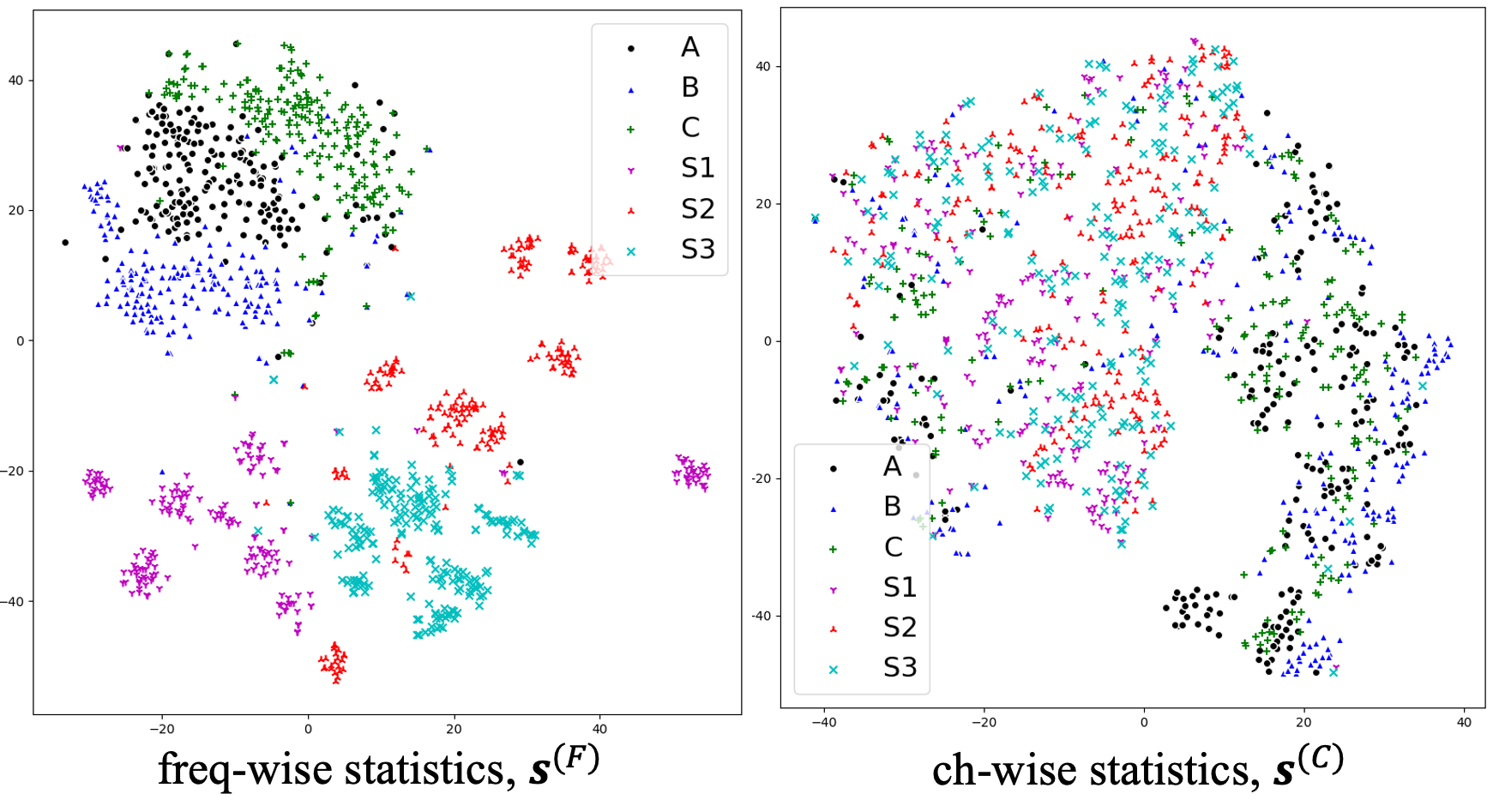}
  \vskip -0.13in
  \caption{\textbf{2D t-SNE visualizations on the DCASE2021 ASC dataset} using activation of stage 1 in BC-ResNet-1.
  }
  \vskip -0.2in
  \label{fig:tsne}
\end{figure}

\noindent \textbf{t-SNE visualization.} We compare $s^{(F)}$ and $s^{(C)}$ through the 2D t-SNE visualization~\cite{tsne} in Figure~\ref{fig:tsne}. The features are separated better by device-IDs (A to S3) with frequency-wise mean and std, $s^{(F)}$, compared to that of the channel, $s^{(C)}$. The observation supports the analysis of MI estimation. 

\section{Relaxed Instance Frequency-wise Normalization}
\label{sec:defineRFN}

Instance Normalization (IN)~\cite{IN} has been one of the promising ways to get domain-invariant features for various domain generalization~\cite{BIN, IBNNet, metabin} and style transfer approaches~\cite{adain} in image processing. However, we showed that domain information is highly contained in the frequency dimension in audio. Therefore, we consider using instance frequency-wise normalization (IFN) instead of IN for the domain generalization in audio.

\noindent \textbf{Frequency-wise normalization.} Before defining IFN, we start by reviewing the formulation of conventional channel-wise normalization (CN), e.g., Batch Normalization (BN)~\cite{BN}, IN, and Group Normalization (GN)~\cite{GN}, to distinguish them from frequency-wise normalization (FN). We denote an element of a normalized feature as $\hat{x}_i=(x_i-\mu_i)/\sigma_i$. Here, $\mu_i = \frac{1}{m}\sum_{k\in S_i}x_k$, $\sigma_i^2 = \frac{1}{m}\sum_{k\in S_i}(x_k-\mu_i)^2+\epsilon$ are calculated over an index-set $S_i$, whose size $|S_i| = m$, and $\epsilon$ is a small constant.
Then, feature normalization methods can be defined by $S_i$. We define CN as feature normalization, where the feature elements with the same channel index, $i_C$, are normalized together, e.g., BN is defined by $S_i = \{k|k_C=i_C\}$. Similarly, we define FN as feature normalization, where feature elements sharing the same frequency index, $i_F$, are normalized together. Then, IFN is defined by
\begin{equation}
\label{eq:IFN}
    \text{IFN: }\; S_i = \{k|k_N=i_N, k_F=i_F\}.
\end{equation}

\noindent \textbf{Relaxed instance frequency-wise normalization.}
We use IFN to eliminate instance-specific domain discrepancy represented in the frequency distribution. However, Section~\ref{sec:2d_audio} showed that frequency statistics are also highly correlated to class-discriminative information. Thus, we try to \textit{relax} (alleviate) the possible loss of useful information by adding an additional feature that is not frequency-wise normalized. We use instance-wise global statistics using $S_i = \{k|k_N=i_N\}$, which is also known as Layer Normalization (LN)~\cite{LN}. Using IFN and LN, we introduce a novel domain generalization module, Relaxed instance Frequency-wise Normalization (RFN), as follows:
\begin{equation}
\label{eq:RFN}
\textit{RFN}(x) = \lambda \cdot \textit{LN}(x) \ + (1-\lambda)\cdot\textit{IFN}(x),\\
\end{equation}
where $x$ is input to RFN, and $\lambda \in [0,1]$ represents the degree of relaxation.
We do not use affine transformation for IFN and LN, and
the resulting mean and std of $\textit{RFN}(x)$ are $\hat{\mu}^{(F)}_i = \frac{\lambda}{\sigma_i}(\mu_i^{(F)} - \mu_i)$ and $\hat{\sigma}^{(F)}_i = \lambda\cdot \frac{\sigma_i^{(F)}}{\sigma_{i}} + (1-\lambda)$, respectively
, where $\mu_i^{(F)}$ and $\sigma_i^{(F)}$ are calculated over $S_i=\{k|k_N = i_N, k_F = i_F\}$, and $\mu_{i}$ and $\sigma_{i}$ are statistics over $S_i=\{k|k_N = i_N\}$. 
We chose LN with IFN for RFN, but we observed that other normalization methods like BN, GN, and even identity connection \cite{resnorm_dcase_challenge} work well with IFN.
We present our proposed RFN in Algorithm~\ref{alg:RFN}, which is simple and easy to implement.
RFN can be inserted in the middle of the existing network 
as an additional plug-and-play module.

\setlength{\textfloatsep}{3pt}
\begin{algorithm}[t]
\caption{RFN, applied to activations $x$ over a mini-batch.}
\label{alg:RFN}
\begin{algorithmic}
\State \textbf{Input:} Activation $x\in R^{N \times C\times F\times T} $, relaxation $\lambda \in [0,1]$.
\State \textbf{Output:} $y = \textit{RFN}(x)$
\State ~~
\State $\text{E}_{\textit{IFN}}[x] \xleftarrow[]{} \frac{1}{C\cdot T}\sum^C_{c}\sum^T_{t}x_{:,c,:,t}$
\State $\text{Var}_{\textit{IFN}}[x] \xleftarrow[]{} \frac{1}{C\cdot T}\sum^C_{c}\sum^T_{t}(x_{:,c,:,t}-\text{E}_{\textit{IFN}}[x])\odot(x_{:,c,:,t}-\text{E}_{\textit{IFN}}[x])$
\State $\textit{IFN}(x)_{:,c,:,t} \xleftarrow[]{} \frac{x_{:,c,:,t} - \text{E}_{\textit{IFN}}[x]}{\sqrt{\text{Var}_{\textit{IFN}}[x] - \epsilon}}$
\Comment{IFN}

\State ~~
\State $\text{E}_{\textit{LN}}[x] \xleftarrow[]{} \frac{1}{C\cdot F \cdot T}\sum^C_{c}\sum^F_{f}\sum^T_{t}x_{:,c,f,t}$
\State $\text{Var}_{\textit{LN}}[x] \xleftarrow[]{} \frac{1}{C\cdot F \cdot T}\sum^C_{c}\sum^F_{f}\sum^T_{t}(x_{:,c,f,t}-\text{E}_{\textit{LN}}[x])\odot(x_{:,c,f,t}-\text{E}_{\textit{LN}}[x])$
\State $\textit{LN}(x)_{:,c,f,t} \xleftarrow[]{} \frac{x_{:,c,f,t} - \text{E}_{\textit{LN}}[x]}{\sqrt{\text{Var}_{\textit{LN}}[x] - \epsilon}}$
\Comment{LN}

\State ~~
\State $y \xleftarrow[]{} \lambda\cdot \textit{LN}(x) + (1-\lambda)\cdot \textit{IFN}(x)$
\Comment{RFN}
\end{algorithmic}
\end{algorithm}

\begin{table*}[t]
    \caption{\textbf{Acoustic Scene Classification.} Top-1 validation accuracy (\%) on TAU Urban Acoustic Scenes 2020 Mobile development dataset (average and standard deviation; averaged over 5 seeds).}
    \vskip -0.1in
    \label{table:ASC_result}
    \centering
    \resizebox{0.8\linewidth}{!}{
    \begin{tabular}{lcccccccc|ccc|c|c}
    \toprule
    \multirow{2}{*}{Method} & \multirow{2}{*}{Backbone} & \multirow{2}{*}{\#Param} & \multicolumn{6}{c}{seen} & \multicolumn{3}{c}{unseen} &  \multirow{2}{*}{Overall} &
    \multirow{2}{*}{$\Delta$}\\
    & & & A & B & C & S1 & S2 & S3 & S4 & S5 & S6 &\\
    
    \midrule
    
    Vanilla & BC-ResNet-8 & 315k & 79.6 & 70.8 & 74.3 & 69.8 & 66.2 & 72.8 & 63.6 & 63.3 & 59.2 & 68.9 $\pm$ 0.8 & + 0.0\\
    Global FreqNorm & BC-ResNet-8 & 315k & 80.2 & 72.2 & \textbf{76.2} & 70.8 & 67.5 & 72.6 & 65.4 & 66.0 & 56.2 & 69.7 $\pm$ 0.6 & + 0.8\\
    PCEN & BC-ResNet-8 & 315k & 75.6 & 66.7 & 66.3 & 69.0 & 67.0 & 73.6 & 68.1 & 68.3 & 66.7 & 69.0 $\pm$ 0.7 & + 0.1 \\
    Mixup & BC-ResNet-8 & 315k & 79.9 & 70.3 & 72.0 & 69.8 & 65.9 & 70.1 & 60.5 & 60.8 & 56.1 & 67.3 $\pm$ 1.0 & - 1.6\\
    MixStyle & BC-ResNet-8 & 315k & 78.5 & 70.0 & 72.0 & 68.4 & 65.9 & 68.3 & 59.0 & 59.3 & 54.6 & 66.2 $\pm$ 0.7 & - 2.7\\
    BIN & BC-ResNet-8 & 317k & 76.9 & 70.2 & 71.4 & 67.3 & 65.6 & 69.6 & 60.4 & 62.2 & 57.6 & 66.8 $\pm$ 1.5 & - 2.1 \\
    CSD & BC-ResNet-8 & 317k & 77.5 & 71.4 & 72.8 & 68.7 & 66.8 & 71.0 & 65.0 & 63.5 & 56.7 & 68.2 $\pm$ 0.4 & - 0.7 \\
    \midrule
    \textbf{RFN (Ours)} & BC-ResNet-8 &  315k & \textbf{82.4} & \textbf{73.2} & 74.5 & \textbf{75.7} & 69.9 & \textbf{76.9} & \textbf{70.5} & \textbf{72.4} & \textbf{69.5} & \textbf{73.9 $\pm$ 0.7} & \textbf{+ 5.0}\\
    ~~ $\lambda=0$ (IFN) & BC-ResNet-8 & 315k & 77.1 & 71.7 & 67.8 & 73.0 & \textbf{71.1} & 74.8 & 69.8 & 70.9 & 67.4 & 71.5 $\pm$ 1.2 & + 2.6\\
    ~~ $\lambda=1$ (LN) & BC-ResNet-8 & 315k & 79.8 & 71.9 & 72.9 & 73.6 & 68.9 & 70.7 & 62.2 & 63.2 & 57.2 & 68.9 $\pm$ 0.7 & + 0.0\\
    \midrule
    \midrule
    Vanilla & BC-ResNet-1 & 8.1k & 73.3 & 61.3 & \textbf{64.9} & 61.0 & 58.3 & 66.7 & 51.8 & 51.3 & 48.5 & 59.7 $\pm$ 1.3 &  + 0.0\\
    \textbf{RFN (Ours)} & BC-ResNet-1 & 8.1k & \textbf{75.2} & \textbf{63.7} & 64.0 & \textbf{62.8} & \textbf{61.2} & \textbf{68.0} & \textbf{58.3} & \textbf{63.0} & \textbf{57.2} & \textbf{63.7 $\pm$ 0.9} & \textbf{+ 4.0} \\
    
    \midrule
    \midrule
    Vanilla & CP-ResNet & 897k & 78.1 & \textbf{71.2} & \textbf{73.4} & 68.3 & 65.9 & 68.7 & 64.8 & 64.8 & 58.5 & 68.2 $\pm$ 0.4 & + 0.0\\
    \textbf{RFN (Ours)} & CP-ResNet & 897k & \textbf{79.3} & 70.9 & 70.8 & \textbf{71.8} & \textbf{72.1} & \textbf{74.1} & \textbf{69.9} & \textbf{68.6} & \textbf{66.0} & \textbf{71.5 $\pm$ 0.3} & \textbf{+ 3.3}\\
    
    \bottomrule
    \end{tabular}
    }
\vskip -0.2in
\end{table*}

\section{Experiments}
\label{sec:exp}

\subsection{Datasets and Experimental Setup}
\label{sec:exp_setup}


\noindent \textbf{Multi-device audio scene classification.} In ASC, it is required to classify an audio segment to one of the given acoustic scene labels. We use the TAU Urban Acoustic Scenes 2020 Mobile development dataset~\cite{dcase_dataset} from DCASE2021, which consists of training and validation data of 13,962 and 2,970 audio segments, respectively. The data are recorded from 12 European cities in 10 different acoustic scenes using three real (A, B, and C) and six simulated devices (S1-S6). The task is challenging due to domain imbalance (10,215 training data samples are recorded by device A.) and unseen domains (devices S4-S6 are unseen during training). Each recording is 10-sec-long, and the sampling rate is 48kHz. We do downsampling by 16kHz and use input features of 256-dimensional log-Mel spectrograms with a window length of 130ms and a frameshift of 30ms. 

Backbones are the ASC version of BC-ResNets~\cite{resnorm_dcase_challenge} and CP-ResNet, c=64~\cite{cpresnet}. During training, we augment data as follows: (1) We use time-shift of $T$ seconds where
$T \sim \textit{Uniform}[-1.5, 1.5]$; (2) We use Specaugment~\cite{specaugment} with two frequency masks and two temporal masks with mask parameters of 40 and 80, respectively, except time warping. We apply Specaugment only for the large model, BC-ResNet-8, and use it after the first RFN at input features. We train each model for 100 epochs using stochastic gradient descent (SGD) optimizer with momentum set to 0.9 and weight decay to 0.001. We use the mini-batch size of 100 and 64 for BC-ResNet-1 and 8, respectively. The learning rate linearly increases from 0 to 0.1 and 0 to 0.06 over the first five epochs as a warmup~\cite{warmup} for BC-ResNet-1 and 8, respectively. Then it decays to zero with cosine annealing\cite{cosine_schedule} for the rest of the training. CP-ResNet is trained in the same manner as BC-ResNet-8. We report the validation performance of each trial after the last epoch. We add RFN at the input and after every \textit{stage} with $\lambda=0.5$ as default.

\noindent \textbf{Baselines.}
The literature in domain generalization (DG) is vast and out of scope for this work. Hence, we compare our approach to the most relevant statistics-based normalization techniques. First, We use (1) `Global FreqNorm' to normalize inputs using global frequency-wise statistics from training data as preprocessing. Next, we experiment (2) Per-Channel Energy Normalization (PCEN)~\cite{pcen}, which preprocess audio input by the moving average of frequency-wise energy along the temporal axis. We use PCEN instead of log-Mel with hyperparameters introduced in \cite{pcen}. Finally, there have been IN-based approaches, e.g., (3) MixStyle~\cite{DG_mixstyle}, mixing channel-wise statistics within mini-batch in the middle of a network, and (4) BIN~\cite{BIN}, a combination of BN and IN, which switches all BNs in a network. In addition, we compare notable approaches from other DG branches, (5) a general regularization approach, Mixup~\cite{mixup}, (6) an adversarial gradient-based method, CrossGrad~\cite{CG}, and (7) a decomposition method, Common-Specific Decomposition (CSD)~\cite{CSD}.
We use the official implementations of MixStyle, BIN, Mixup, and CSD~\cite{DG_mixstyle, BIN, mixup, CSD} and follow their settings.

\subsection{Experimental Results}
\label{sec:exp_results}

\textbf{Multi-device acoustic scene classification.} 
The results are shown in Table~\ref{table:ASC_result}. The vanilla BC-ResNet-8 shows poor device generalization capability. The dominant one, `A,' shows the highest performance while the unseen devices' performance is much lower compared to seen devices. Overall, the baselines do not show clear improvements compared to the vanilla BC-ResNet-8 for both seen and unseen domains. PCEN seems to help device generalization by frequency-wise input preprocessing, but it leads to performance degradation for seen devices. 
Our RFN shows significant improvements, more than 10 \% in top-1 accuracy for unseen domains, while still showing promising results for seen devices. However, when we directly use IFN by $\lambda=0$, it shows good generalization capability but degrades the performance for seen devices compared to RFN. Also, the direct use of LN by $\lambda=1$ does not show a clear improvement. 

We experiment with a severe case using only `A, S1, S2, and S3' for training. For additional unseen devices `B' and `C', vanilla BC-ResNet-8 gets 41.4 \% and 54.2 \%, respectively, and RFN gets improved 49.2 \% and 62.0 \% accuracies, respectively.


\begin{table}[t]
\caption{\textbf{Top-1 accuracy (\%) of frequency-wise approaches and their channel-wise counterparts} (averaged over 5 seeds).}
    \vskip -0.05in
    \label{compare_ch_freq}
    \centering
    \resizebox{0.7\linewidth}{!}{
    \begin{tabular}{lcc|c}
    \toprule
    Method & seen & unseen & Overall \\
    \midrule
    Baseline & 72.3 & 62.0 & 68.9 $\pm$ 0.8\\
    \midrule
    Global Norm & 71.8 & 60.3 & 68.0 $\pm$ 0.7 \\
    BIN & 70.2 & 60.1 & 66.8 $\pm$ 1.5\\
    MixStyle & 70.5 & 57.6 & 66.2 $\pm$ 0.7 \\
    Relaxed-IN & 72.9 & 60.7 & 68.8 $\pm$ 0.6 \\
    \midrule
    Global FreqNorm & 73.2 & 62.5 & 69.7 $\pm$ 0.6 \\
    BIFN & 72.0 & 62.6 & 68.8 $\pm$ 1.0 \\
    Freq-MixStyle & 73.2 & 67.7 & 71.3 $\pm$ 0.9 \\
    \textbf{RFN (Ours)} & \textbf{75.4} & \textbf{70.8} & \textbf{73.9 $\pm$ 0.7} \\
    
    
    \bottomrule
    \end{tabular}
    }
\vskip -0.1in
\end{table}

\subsection{Ablation Studies}
\label{sec:ablations}

\textbf{Importance of frequency-wise normalization.} We compare various channel-wise approaches to their frequency-wise counterparts in Table~\ref{compare_ch_freq}. `Global Norm' normalizes inputs using channel-wise global statistics over training dataset, and its frequency-wise version is `Global FreqNorm.' 
We use IFN instead of IN to get `BIFN' from BIN, and `Freq-MixStyle' mixes frequency-wise statistics rather than channel-wise statistics. We apply Freq-MixStyle at the input and the end of every stage in BC-ResNet-8 like RFN. We also use `Relaxed-IN' as a counterpart of RFN. The results show consistent improvements for frequency-wise approaches compared to each of their channel-wise version.
Thus, the results support the importance of frequency-wise approaches in audio, and RFN still outperforms other frequency-wise versions of the baselines.

\begin{figure}[t]
\centering
  \includegraphics[width=0.7\linewidth]{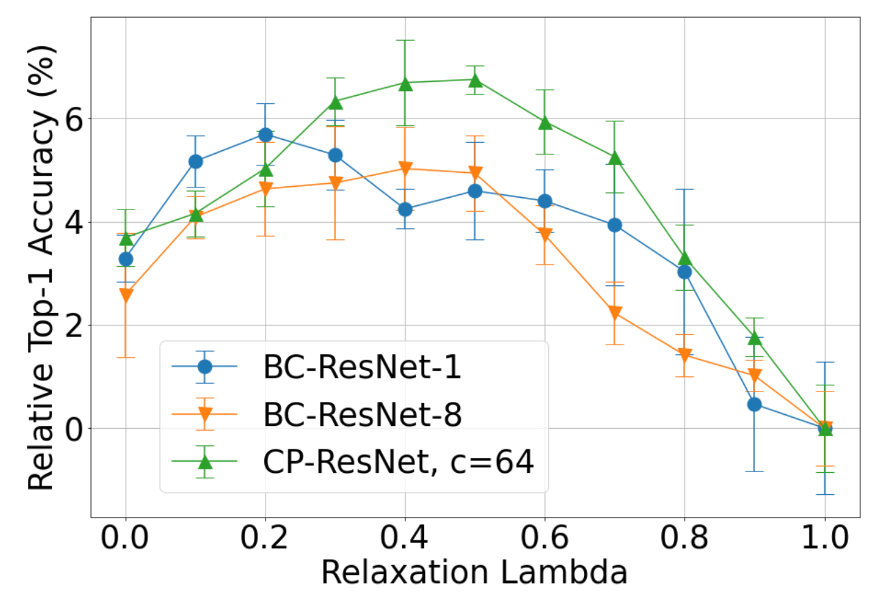}
  \vskip -0.1in
  \caption{\textbf{Relative Acc. changes} as $\lambda$ changes compared to $\lambda=1$ (Average over 5 seeds; error bar stands for std).
  }
  \label{fig:lambda}
\end{figure}

\noindent \textbf{Sensitivity to the degree of relaxation.}
Figure~\ref{fig:lambda} shows how the varying $\lambda$ affects the performance. For clarity, we draw the graphs of relative performance compared to $\lambda=1$ for the experiments of Section~\ref{sec:exp_results}. 
The optimal $\lambda$ is between 0 and 1 for various architectures, which implies the necessity of relaxation.

\begin{table}[t]
\caption{\textbf{Keyword Spotting.} Top-1 test accuracy (\%) with varying number of training speakers on Google speech command dataset ver1 (average and std; averaged over 5 seeds).
}
\vskip -0.25in
\label{table:kws_dg}
\begin{center}
    \resizebox{0.9\linewidth}{!}{
    \begin{tabular}{l|c|c|c|c}
    \toprule
    Method & 50 & 100 & 200 & 1000 \\
    \midrule
    cnn-trad-fpool3 & 72.5 $\pm$ 0.3 & 80.5 $\pm$ 0.5 & 86.9 $\pm$ 0.5 & 92.3 $\pm$ 0.4 \\
    ~~ + Mixup & 70.5 $\pm$ 0.5 & 79.7 $\pm$ 0.4 & \textbf{87.9 $\pm$ 0.4} & \textbf{93.3 $\pm$ 0.4} \\
    ~~ + CrossGrad & 73.7 $\pm$ 0.8 & 81.1 $\pm$ 0.5 & 87.2 $\pm$ 0.2 & 92.6 $\pm$ 0.2 \\
    ~~ + CSD & 73.4 $\pm$ 0.5 & 80.9 $\pm$ 0.6 & 87.5 $\pm$ 0.5 & 92.8 $\pm$ 0.3 \\
    
    
    \midrule
    
    ~~ + \textbf{RFN (Ours)} & \textbf{76.6 $\pm$ 0.6} & \textbf{82.3 $\pm$ 0.6} & \textbf{87.8 $\pm$ 0.6} & 92.4 $\pm$ 0.3 \\
    ~~ ~~ $\lambda=0$ (IFN) & 71.1 $\pm$ 1.1 & 77.9 $\pm$ 0.7 & 85.3 $\pm$ 0.3 & 90.9 $\pm$ 0.3\\
    ~~ ~~ $\lambda=1$ (LN) & 72.7 $\pm$ 0.8 & 80.8 $\pm$ 0.7 & 86.5 $\pm$ 0.3 & 92.4 $\pm$ 0.2 \\
    

    \bottomrule
    \end{tabular}
    }
\end{center}
\vskip -0.25in
\end{table}

\noindent \textbf{Apply RFN to other tasks.} Finally, we apply RFN to other two tasks, (1) keyword spotting and (2) speaker verification.
\textbf{(1)} We follow the benchmark settings introduced in \cite{CSD}, which experiment with a backbone architecture, \textit{cnn-trad-fpool3}~\cite{cnn_kws_interspeech15}, with a varying number of training speakers from 50 to 1,000 on the Google speech command dataset~\cite{google_commands} (keyword spotting). 
Here, the task desires robustness over speaker IDs. 
We use RFN only at the input as an additional module, considering the shallow architecture of the backbone. In Table~\ref{table:kws_dg}, compared to baselines, RFN especially shows clear margins when the number of training speakers is small. 
\textbf{(2)} We further experiment RFN to multi-genre speaker verification task using CN-Celeb dataset~\cite{cnceleb1, cnceleb2} based on the settings of~\cite{fastresnet1}.
The task requires robustness over 11 genres, e.g., `interview,' `drama,' and `singing.' We exploit Fast-ResNet-34~\cite{fastresnet1}, suitable for speaker verification, and train the model with AM-Softmax~\cite{AMsoftmax1}.
The `baseline' uses a conventional input normalization method in SV, global mean subtraction. We measure equal error rates (EER) (lower is better) on both the overall validation set and each genre validation set except `advertisement,' `play,' and `recitation,' which consists of a very few positive pairs following the settings of \cite{cnceleb2} and \cite{MAML_SV}. We show the results in Table~\ref{table:sv}.
The conventional normalization method produces poor generalization for novel genres, {\it i.e.}, `movie' and `singing.' RFN is also helpful for this metric learning task and brings a better overall EER of 13.4 \%.


\begin{table}[t]
\caption{\textbf{Speaker verification.} EER (\%) on overall genres and each genre on CN-Celeb. The numbers are average and standard deviation (average over 5 seeds).}
\label{table:sv}
    \vskip -0.1in
    \centering
    \resizebox{1.0\linewidth}{!}{
    \begin{tabular}{l|cccccc|cc|c}
    \toprule
    \multirow{2}{*}{Method} & \multicolumn{6}{c}{seen} & \multicolumn{2}{c}{unseen} &  \multirow{2}{*}{Overall} \\ 
    & drama & entertainment & interview & live & speech & vlog & movie & singing & \\
    \midrule
    Baseline & 13.1 & 14.6 & 11.4 & 13.8 & 5.7 & 7.6 & 18.4 & 28.7 & 14.5 $\pm$ 0.4 \\
    \midrule
    Mixup & 13.9 & 14.3 & 10.7 & \textbf{12.0} & 4.6 & \textbf{6.3} & \textbf{16.8} & 28.9 & 14.1 $\pm$ 0.2\\
    MixStyle & 14.2 & 14.2 & 11.1 & 13.0 & 5.5 & 7.0 & 19.7 & 28.1 & 14.0 $\pm$ 0.3 \\
    BIN & 14.1 & 15.5 & 12.2 & 14.0 & 5.3&  8.4& 18.3 & 29.5 & 15.0 $\pm$ 0.6 \\
    \midrule
    \textbf{RFN (Ours)} & \textbf{11.9} & \textbf{13.5} & \textbf{10.1} & 12.9& \textbf{4.1} & 7.2& 18.7& \textbf{27.4} &\textbf{ 13.4 $\pm$ 0.3} \\ 
    \bottomrule
    \end{tabular}
    }
\end{table}

\section{Conclusion}
\label{sec:conclusion}
We address the characteristics of audio features in 2D-CNNs and show a guideline to get audio domain invariant features. Frequency-wise distribution is highly correlated to domain information, and we can eliminate instance-specific domain discrepancy by explicitly manipulating frequency-wise statistics rather than channel statistics. Based on the analysis, we introduce a domain generalization method, Relaxed instance Frequency-wise Normalization (RFN).


\bibliographystyle{IEEEtran}

\bibliography{mybib}


\end{document}